\documentclass[aps,showpacs]{revtex4}
\usepackage{epsfig,amsmath,amsfonts,wasysym}
\usepackage{dsfont}
\usepackage{graphicx}
\newcommand{\Ref}[1]{(\ref{#1})}
\newcommand{\adb}{\allowdisplaybreaks }
\newcommand{\TE}{{\rm TE}}
\newcommand{\TM}{{\rm TM}}
\newcommand{\te}{\textsf{\scriptsize TE} }
\newcommand{\tm}{\textsf{\scriptsize TM} }
\newcommand{\bs}{\begin{subequations}}
\newcommand{\es}{\end{subequations}}
\begin{document}

\title{Van der Waals interaction between an atom with spherical plasma shell}
\author{Nail R. Khusnutdinov\footnote{e-mail: 7nail7@gmail.com}}

\affiliation{Institute of Physics, Kazan Federal University, Kremlevskaya 18,
Kazan, 420008, Russia}

\begin{abstract}
The van der Waals interaction energy of an atom with infinitely thin sphere with finite conductivity is investigated in the framework of the hydrodynamic approach. Thin sphere models the fullerene. We put the sphere into spherical cavity inside the infinite dielectric media, then calculate the energy of vacuum fluctuations in the context of the zeta-function approach. The interaction energy for a single atom is obtained from this expression in the limit of the rare media. The Casimir-Polder expression for an atom and plate is recovered in the limit of the infinite radius of the sphere. Assuming a finite radius of the sphere, the interaction energy of an atom falls down monotonic as third power of distance between atom and sphere for short distance and as seventh power for large distance from the sphere. Numerically, the interaction energy is obtained to be $3.8 eV$ for hydrogen atom placed on the surface of the sphere with parameters of fullerene $C_{60}$. We show also that the polarizability of fullerene is merely cube of its radius.        
\end{abstract}  

\pacs{73.22.-f, 34.50.Dy, 12.20.Ds}
%\keywords{zeta-function;
%Casimir effect; Zero-point energy; Renormalization; Heat-kernel
%coefficients}98.80.Cq, 14.80.Hv

\maketitle

\section{Introduction}\label{Sec:Intro}
The general theory of the van der Waals force was developed by Lifshits in Refs.
\cite{Lif56,LifPit80} in the framework of  statistical physics. In the case of
interaction between particle and plate it is commonly referred to as the
Casimir-Polder force \cite{CasPol48}. For small distance the potential of
interaction is proportional to inverse third degree of distance from the plate.
For large distance the retardation of the interaction is taken into account and
the potential falls down as fourth degree of distance. The last achievements in
Casimir effect have been discussed in great depth in books and reviews  
\cite{Mil01,BorMohMos01,BorKliMohMos09,KliMohMos09}.

The van der Waals force is very important for interaction of graphene
(graphite layers) with bodies
\cite{BogOveHylLunBruJen00,HulHylLun01,RydDioJacShrHylSimLanLun03,%
JunGarDobGod04,KleiHylSch05,DobWhiRub06,BorGeyKliMos06,BorFiaGitVas09} and
microparticles \cite{DinNakKas04,BonLam04,BonLam05,BlaKliMos07,ChuFedKliYur10}.
An understanding of the mechanisms of molecule-nanostructure interaction is of
importance for the problem of hydrogen storage in carbon nanostructures
\cite{DilJonBekKiaBetHeb97}. The microscopic mechanisms underlying the
absorption phenomenon remain unclear (see, for example review \cite{Nec06}).

In the present paper we use model of the fullerene in terms of the two
dimensional free electron gas \cite{Fet73} which is usually called as
hydrodynamical model.  This model was applied and developed for the molecule
$C_{60}$ in Refs. \cite{Bar04,Bar05}, for flat plasma sheet  in Ref.
\cite{BorPirNes05} and for spherical plasma surface in Ref. \cite{BorKhu08}. In
the framework of this model the conductive surface is considered as infinitely
thin shell with the specific wave number $\Omega = 4\pi n e^2/mc^2$, where $n$
is surface density of electrons and $m$ is the electron mass. Since the surface is infinitely thin, the information about the properties
of the surface is encoded in the boundary conditions on the conductive surface
which are different for TE and TM modes. In the Ref. \cite{BorKhu08} it was shown that 
the energy of the vacuum electromagnetic fluctuations for surface shaped as
sphere has a maximum for radius of sphere approximately equal to the specific
wavelength of the model $\lambda_\Omega = 2\pi/\Omega$. What this means is the
Casimir force tries to enlarge sphere with radius larger then $\lambda_\Omega$
and it tries to reduce the sphere with radius larger then $\lambda_\Omega$. The Boyer result \cite{Boy68} is recovered in the limit $\Omega \to \infty$. 

At the same time it is well known \cite{GeiNov07} that the energy of electrons
in graphene has linear frequency dependence whereas in framework of the
hydrodynamic model the energy of electrons is quadratic in the frequency. There 
is also another point that the electrons in the graphene have zero or very small
effective mass. To describe correctly these unusual properties of electrons in
graphene the Dirac fermion model was suggested in Ref. \cite{Sem84}. The
electrons in this model are described by $(2+1)D$ Dirac action with
characteristic propagation velocity as Fermi velocity $v_F \approx c/300$ and
very small mass gap $m < 0.1 eV$. This model was applied for calculation of
Casimir interaction energy between graphene plate and perfect conductor plane
in Ref. \cite{BorFiaGitVas09} and recently in Ref. \cite{ChuFedKliYur10} for
Casimir-Polder interaction energy between graphene and H, He$^*$ and Na atoms.

It was shown that the Casimir energy for large distance between graphene plate
and perfect conductor plane \cite{BorFiaGitVas09} is decreasing by one power of
the separation a faster than for ideal conductors, that is as $(ma)^{-4}$. If
the mass of gap is zero at the beginning of calculations, $m=0$, they obtained
standard dependence $a^{-3}$. For the case  of Casimir-Polder interaction
energy between graphene and atoms \cite{ChuFedKliYur10} the hydrodynamic and
the Dirac models give qualitatively different results. For the large separation the energy decreases with separation as $a^{-4}$ which is a typical behavior of the atom-plate interaction at relativistic separations, but the coefficients are different. In the case of H, He$^*$ and Na atoms, the hydrodynamic model gives $\approx 5$ times larger coefficient than the Dirac model. There is also interesting observation about mass gap parameter: the energy does not depend on the parameter for $m < 10^{-3/2} eV$ and therefore the limit $m\to 0$ is satisfied.    

There is another approach for van der Waals interaction based on the density-functional theory \cite{HohKoh64,KohSha65} and the local-density approximation \cite{KohSha65} which has proved to be a very useful tool for calculating the ground-state properties of atoms, molecules, and solids. In framework of density-functional theory a number of studies of van der Waals interaction has been made \cite{RapAsh91,AndLanLun96,HulAndLun96,DobDin96,KohMaiMak98,DobWan00,%
RydDioJacSchHylSimLanLun03,DioRydSchLanLun04,DobWhiRub06,LeeMurKonLunLan10}. The main problem in this theory is to find approximations for the exchange-correlation energy. The density-functional theory describes cohesion, bonds, structures, and other properties very well for dense molecules and materials. The theory fails to describe the
interactions at sparse electron densities. The solution of this point by introducing  the non-local correlations may be found in Refs.  \cite{RydDioJacSchHylSimLanLun03,DioRydSchLanLun04,LeeMurKonLunLan10}.

In the present paper the hydrodynamical model of fullerene is adopted -- the  infinitely thin sphere with radius $R$ in vacuum and finite conductivity. To obtain the van der Waals interaction energy between an atom and this sphere we use the following approach which is due to Lifshits (see Refs. \cite{Lif56,LifPit80,BorGeyKliMos06,BlaKliMos07}). We put the sphere inside the spherical vacuum cavity with radius $L = R+d > R$ which is inside the dielectric media with coefficients $\mu, \varepsilon$. Then we find the zero-point energy of this system by using the zeta-function regularization approach, and take the limit of the rared media with $\varepsilon = 1+ 4\pi N \alpha + O(N^2)$, where $N\to 0 $
is the volume density of the atoms and $\alpha$ is the polarizability of the
unit atom. The interaction energy per unit atom which is situated $d$ from the sphere is found by simple formula 
\begin{equation*}
E_a(s) = -\lim_{N\to 0}\frac{\partial_d E(s)}{4\pi N (R+d)^2},
\end{equation*}
where $E(s)$ is the zeta-regularized energy with regularization parameter $s$.  

The paper is organized as follows. In Sec. \ref{Sec:Maxwell} we derive the boundary conditions for electromagnetic field on the infinitely thin conductive sphere as well as on the boundary of the cavity. Section \ref{Sec:Matching} is devoted to the construction of solutions satisfying the boundary conditions. The expression for the van der Waals energy is found in Sec. \ref{Sec:Energy} and it is analyzed in the limits of infinite radius of the sphere and for short and large distances between atom and sphere. Section \ref{Sec:Num} contains the numerical calculations of the interaction energy between hydrogen atom and the infinitely thin sphere with parameters of the fullerene $C_{60}$. In the last section \ref{Sec:Conc} we discuss results obtained.

\section{Maxwell's equations and matching conditions}\label{Sec:Maxwell}

\begin{figure}[ht]
%\centerline{\epsfxsize=5truecm\epsfbox{1.eps}}
\includegraphics[width=5truecm]{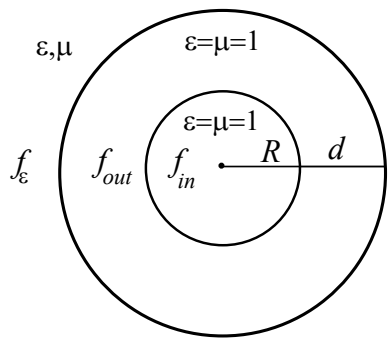}
\caption{\footnotesize The infinitely thin conductive sphere with radius $R$ is located inside the vacuum spherical cavity with radius $L=R+d$ into dielectric media with $\varepsilon,\mu \not = 1$.}\label{fig:spheres}
\end{figure}

Let us consider a conductive infinitely thin sphere with radius $R$ in vacuum spherical cavity with radius $L=R+d$ which is inside the dielectric media with parameters $\mu, \varepsilon$ (see fig. \ref{fig:spheres}). 
We have two concentric spheres and we should consider the boundary conditions on two spherical boundaries.  

I. First of all let us consider a spherical boundary with radius $L=R+d$. Inside the sphere we have vacuum, $\varepsilon = \mu = 1$ and outside -- the dielectric media with $\varepsilon, \mu \not = 1$. Assuming the spherical symmetry, the  electromagnetic field is factorized for two independent polarizations usually called as \te and \tm modes. The Maxwell equations with oscillatory time dependence $\exp (-i\omega t)$ read
\bs\label{Max1}
\begin{eqnarray}
 \textrm{rot \!} \mathbf{E} - \frac{i\omega}c \mathbf{B} &=& 0,\ \textrm{div \!} \mathbf{B} = 0,\adb \label{Max1-1} \\
 \textrm{rot \!} \mathbf{H} + \frac{i\omega}c \mathbf{D} &=& 0,\ \textrm{div \!} \mathbf{D} = 0, \label{Max1-2}
\end{eqnarray}
\es
where we should use the material equations $\mathbf{D} = \varepsilon (\omega) \mathbf{E}$ and $\mathbf{B} = \mu (\omega) \mathbf{H}$. For \te mode to be obtained we express $\mathbf{B}$ from the first equation (\ref{Max1-1}) and substitute it into the second Eq. (\ref{Max1-2})
\begin{equation}
 \mathbf{B}^\te = -\frac{i c}{\omega} \textrm{rot \!} \mathbf{E}^\te,\ \triangle \mathbf{E}^\te - \frac{\omega^2}{c^2} \mu\varepsilon \mathbf{E}^\te = 0. 
\end{equation}
For \tm mode to be obtained we express $\mathbf{E}$ from second equation (\ref{Max1-2}) and substitute it into the first Eq. (\ref{Max1-1})
\begin{equation}
 \mathbf{E}^\tm = \frac{ic}{\omega \mu\varepsilon} \textrm{rot \!} \mathbf{B}^\tm, \  \triangle \mathbf{B}^\tm - \frac{\omega^2}{c^2} \mu\varepsilon \mathbf{B}^\tm = 0.   
\end{equation}
We next expand solutions over spherical functions, $Y_{lm}$, and obtain the following expressions for \te and \tm polarizations, 
\begin{eqnarray}
 \mathbf{B}_{lm}^\te &=& -\frac{ic}{\omega} \textrm{rot \!} \mathbf{E}_{lm}^\te, \  \mathbf{E}_{lm}^\te = f (kr) \mathbf{L} Y_{lm}, \adb \\ 
 \mathbf{E}_{lm}^\tm &=& \frac{ic}{\omega\mu\varepsilon} \textrm{rot \!} \mathbf{B}_{lm}^\tm, \ \mathbf{B}_{lm}^\tm = f(kr) \mathbf{L}Y_{lm},
\end{eqnarray}
where $ck=\omega \sqrt{\mu\varepsilon}$. In the standard spherical vector basis,    $(e_{r},e_{\theta}, e_{\varphi})$, we obtain in manifest form the modes we need
\begin{eqnarray}
\mathbf{E}_{lm}^\te &=& \left(0,\frac{if}{\sin\theta} \partial_\varphi
Y_{lm}, -if \partial_\theta Y_{lm}\right),\adb \nonumber\\
\mathbf{B}_{lm}^\te &=& \left(\frac{cf}{\omega r} l(l+1)
Y_{lm},\frac{c(rf)'}{\omega r} \partial_\theta Y_{lm},
\frac{c(rf)'}{\omega r \sin\theta} \partial_\varphi Y_{lm}\right), \adb \nonumber\\
\mathbf{B}_{lm}^\tm &=& \left(0,\frac{if}{\sin\theta} \partial_\varphi
Y_{lm}, -if \partial_\theta Y_{lm}\right),\adb \nonumber\\
\mathbf{E}_{lm}^\tm &=& -\frac{c}{\varepsilon\mu}\left(\frac{f}{\omega r}
l(l+1) Y_{lm},\frac{(rf)'}{\omega r} \partial_\theta Y_{lm},
\frac{(rf)'}{\omega r \sin\theta} \partial_\varphi Y_{lm}\right),\label{fields}
\end{eqnarray}
where the function $f$ obeys the following radial equation 
\begin{equation}
f'' + \frac 2r f' + \left( \frac{\omega^2}{c^2}\varepsilon \mu  - \frac{l(l+1)}{r^2}\right)f = 0.
\end{equation}
The two independent solutions of this equation are the spherical Bessel functions $j_l(z) = \sqrt{\pi/2z} J_{l+1/2}(z), \ y_l(z) = \sqrt{\pi/2z} Y_{l+1/2}(z)$, where $z = r \omega \sqrt{\varepsilon\mu}/c$.

At the boundary, $r=L$, the matching conditions read
\bs\label{matching_cond}
\begin{eqnarray}
 \mathbf{n} \cdot [\mathbf{B}_2 - \mathbf{B}_1]_L &=& 0, \ \mathbf{n} \cdot [\mathbf{D}_2 - \mathbf{D}_1]_L =0,\adb\\
 \mathbf{n} \times [\mathbf{H}_2 - \mathbf{H}_1]_L &=& 0, \ \mathbf{n} \times [\mathbf{E}_2 - \mathbf{E}_1]_L =0,
\end{eqnarray}
\es
where $\mathbf{n} = \mathbf{r}/r$ is an unit normal to the sphere. We have to take into account also that $k=\omega /c$ inside the sphere $r=L$ and $k= \omega \sqrt{\mu\varepsilon}/c$ outside the sphere. The square brackets above denote the coincidence limit on the boundary $r=L$.

II. The electromagnetic fields given infinitely thin conductive surface $\Sigma$ in vacuum was considered by Fetter in Ref. \cite{Fet73}. The applications of this model for vacuum fluctuations of field see in Refs. \cite{Bar04,Bar05,BorPirNes05,BorKhu08}. The electrons of conductivity on the sphere produce current and the Maxwell equations read
\bs\label{Max2}
\begin{eqnarray}
 \textrm{rot \!} \mathbf{E} - \frac{i\omega}c \mathbf{H} &=& 0,\ \textrm{div \!} \mathbf{H} = 0,\adb \\
 \textrm{rot \!} \mathbf{H} + \frac{i\omega}c \mathbf{E} &=& 4\pi \mathbf{J},\ \textrm{div \!} \mathbf{E} = 4\pi \rho,
\end{eqnarray}
\es
where $\rho = \delta (\mathbf{x}-\mathbf{x}_\Sigma)\sigma$, $ \mathbf{J} = \delta (\mathbf{x}-\mathbf{x}_\Sigma) \mathbf{j}/c$. Taking into account the equation of continuity and the Newton equations we obtain the following expressions for density and of charge and current on the boundary:
\begin{equation}
\sigma = \frac{e^2 n}{m\omega^2} \nabla_{|\!|} \cdot \mathbf{E}_{|\!|},\ \mathbf{j} = i \frac{e^2 n}{m\omega} \mathbf{E}_{|\!|},
\end{equation}  
where the superscripts $|\!|$ indicate the vector components parallel to the surface $\Sigma$; $e$ and $m$ are the charge and mass of electron, and $n$ is a surface density of charge. 

As a consequence of the charge and current obtained above, the  boundary conditions on the sphere with $r=R$ read
\bs \label{cond2} 
\begin{eqnarray}
\mathbf{n}\cdot [\mathbf{H}_{2} - \mathbf{H}_{1}]_R &=& 0,\ \mathbf{n}\cdot [\mathbf{E}_{2} - \mathbf{E}_{1}]_R = \frac{\Omega }{k^2} \nabla_{|\!|} \cdot \mathbf{E}_{|\!|}, \\
\mathbf{n} \times [\mathbf{H}_2 - \mathbf{H}_1]_R &=& -\frac{i\Omega }{k } \mathbf{n} \times \mathbf{E}_{|\!|},\ \mathbf{n} \times [\mathbf{E}_2 - \mathbf{E}_1]_R = 0,
\end{eqnarray}
\es
where $k=\omega /c$ and $\Omega = 4\pi n e^2/mc^2$ is a specific wave number on the sphere. Because of the fact that the sphere is infinitely thin we may consider the Maxwell equations (\ref{Max2}) in vacuum with zero right hand side and all information about sphere will be encoded in boundary conditions (\ref{cond2}). An interesting treatment of this boundary condition is in Ref. \cite{Vas09}.

\section{The solution of the matching conditions}\label{Sec:Matching}

Let us represent the radial function in the following way
\begin{equation}
f = \left\{ 
 \begin{array}{lll} f_{in}&=a_{in} j_l(kr),& r<R \\
               f_{out}&=a_{out} j_l(kr) + b_{out} y_{l}(kr),&R<r<L \\
               f_{\varepsilon}&=a_{\varepsilon} h^{(1)}_l(kr),&r>L     
 \end{array}\right. 
\end{equation}
where $j_l,y_l$ and $h_l^{(1)}$ are the spherical Bessel functions and $k=\omega /c$ inside the sphere, $r < L$ and $k= \omega \sqrt{\mu\varepsilon}/c$ outside the sphere for $r > L$. 

In this case the matching conditions (\ref{matching_cond}) and (\ref{cond2}) in manifest form read 
\begin{eqnarray}
 [rf_{out} - rf_{in}]_{R} &=&0,\nonumber\adb\\{}
[(rf_{out})'_r -  (rf_{in})'_r - \Omega (rf_{in})]_R &=&0,\nonumber\adb\\{}
[rf_{out} - rf_{\varepsilon}]_{L} &=&0,\adb\\{}
[(rf_{out})'_r -  \frac{1}{\mu}(rf_{\varepsilon})'_r]_L &=&0, \nonumber
\end{eqnarray}
for \te mode, and 
\begin{eqnarray}
 [(rf_{out})'_r - (rf_{in})'_r]_{R} &=&0,\nonumber\adb\\{}
[(rf_{out}) -  (rf_{in}) + \frac{\Omega}{k^2} (rf_{in})'_r]_R &=&0,\nonumber\adb\\{}
[rf_{out} -  \frac{1}{\mu}rf_{\varepsilon}]_L&=&0,\adb\\{}
[(rf_{out})'_r -  \frac{1}{\mu\varepsilon}(rf_{\varepsilon})'_r]_L &=&0,\nonumber
\end{eqnarray}
for \tm mode. The solutions of these equations exist if and only if the following equations are satisfied 
\bs
\begin{eqnarray}
 \frac{1}{\sqrt{\mu\varepsilon}} H(z_\varepsilon) \Psi'_{\te} - \frac{1}{\mu}
H'(z_\varepsilon) \Psi_{\te} &=& 0,\adb\\
 -\frac{1}{\sqrt{\mu\varepsilon}} H(z_\varepsilon) \Psi'_{\tm} +
\frac{1}{\varepsilon} H'(z_\varepsilon) \Psi_{\tm} &=& 0,
\end{eqnarray}
\es
where $z_\varepsilon = z \sqrt{\mu\varepsilon}$, $z = kL = \omega L/c$; the prime is derivative with respect the argument, and
\bs
\begin{eqnarray}
 \Psi_{\te}(z) &=& J(z) + \frac{\Omega}{k} J(x) [J(x) Y(z) - J(z) Y(x)],\adb\\
 \Psi_{\tm}(z) &=& J(z) + \frac{\Omega}{k} J'(x) [J'(x) Y(z) - J(z) Y'(x)].
\end{eqnarray}
\es
Here $J(x) = xj_l(x),\ Y(x) = xy_l(x),\ H(x) = xh^{(1)}_l(x)$ are the Riccati-Bessel functions, and $x=kR$. 
Therefore the functions we need (see next section) to obtain the spectrum of the energy read (we set $\mu =1$) 
\bs
\begin{eqnarray}
 \Sigma_{\te} &=& H'(z_\varepsilon) \Psi_{\te} -
\frac{1}{\sqrt{\varepsilon}} H(z_\varepsilon) \Psi'_{\te} ,\adb\\
 \Sigma_{\tm} &=& H(z_\varepsilon) \Psi'_{TM} -
\frac{1}{\sqrt{\varepsilon}} H'(z_\varepsilon) \Psi_{\tm}.
\end{eqnarray}
\es
For $\varepsilon =1$, the result obtained in the Ref. \cite{BorKhu08} is recovered  
\bs
\begin{eqnarray}
 \Sigma_{\te} &=& i\left\{1 - \frac{\Omega}{ik} J(x)H(x) \right\}=if_{\te}(k),\adb\\
 \Sigma_{\tm} &=& -i\left\{1 - \frac{\Omega}{ik} J'(x)H'(x) \right\} = -
if_{\tm}(k),
\end{eqnarray}
\es
for real value of $k$, and for imaginary axis $k\to ik$ we obtain from above expressions the Jost functions in imaginary axis:
\bs
\begin{eqnarray}
 \Sigma_{\te} &=& i\left\{1 + \frac{\Omega}{k} s_l(x)e_l(x) \right\} = i
f_{\te}(ik),\adb\\
 \Sigma_{\tm} &=& -i\left\{1 - \frac{\Omega}{k} s'_l(x)e'_l(x) \right\} =
-i f_{\tm}(ik),
\end{eqnarray}
\es
because $H(ix) = (-i)^{l+1} e_l(x)$, $J(ix) = i^{l+1} s_l(x)$ and $Y(ix) =
-i^{l} s_l(x) -(-i)^{l} e_l(x)$, where 
\begin{equation}
 s_l(x) = \sqrt{\frac{\pi x}{2}} I_{l+1/2}(x), \ e_l(x) = \sqrt{\frac{2 x}{\pi}}
K_{l+1/2}(x)
\end{equation}
are the Riccatti-Bessel spherical functions of the second kind. For the problem with $z=0$ to be avoided we multiply $\Sigma_{\tm}$ for $z^2$
\bs
\begin{eqnarray}
 \Sigma_{\te} &=& -i\left\{ H'(z_\varepsilon) \Psi_{\te} -
\frac{1}{\sqrt{\varepsilon}} H(z_\varepsilon) \Psi'_{\te}\right\} ,\adb\\
 \Sigma_{\tm} &=& -iz^2\left\{H(z_\varepsilon) \Psi'_{\tm} -
\frac{1}{\sqrt{\varepsilon}} H'(z_\varepsilon) \Psi_{\tm}\right\}.
\end{eqnarray}
\es

On the imaginary axis $k\to i k$ we obtain 
\bs \label{sigmas}
\begin{eqnarray}
 \Sigma_{\te} &=& \frac{1}{\sqrt{\varepsilon}} e_l(z_\varepsilon)
\Phi'_{\te} - e'_l(z_\varepsilon) \Phi_{\te}  ,\adb\\
 \Sigma_{\tm} &=& z^2\left\{e_l(z_\varepsilon) \Phi'_{\tm} -
\frac{1}{\sqrt{\varepsilon}} e_l'(z_\varepsilon) \Phi_{\tm}\right\},\adb\\
 \Phi_{\te} &=& s_l(z) + \frac{Q}{x} s_l(x) [s_l(z) e_l(x) - s_l(x) e_l(z)],\adb\\
 \Phi_{\tm} &=& s_l(z) - \frac{Q}{x} s'_l(x) [s_l(z) e'_l(x) - s'_l(x) e_l(z)],
\end{eqnarray}
\es
where $Q=\Omega R$, $z = kL,\ z_\varepsilon = z \sqrt{\varepsilon},\ x = kR$ and
$\varepsilon = \varepsilon (i\omega)$. For $\varepsilon = 1$ we obtain
\begin{equation}
 \Sigma_{\te} = f_{\te}(ik), \ \Sigma_{\tm} = z^2f_{\tm}(ik)
\end{equation}
in accordance with Ref. \cite{BorKhu08}.

\section{The energy}\label{Sec:Energy}
Within the limits of approach suggested in Ref. \cite{BorEliKirLes97}, the expressions for \te and \tm contributions in regularized  zero-point energy read ($\omega=kc$)
\begin{eqnarray}
E^{\te}(s) &=& -\frac{\hbar c\cos\pi s}{\pi} \mu^{2s}\sum_{l=1}^\infty \nu \int_0^\infty dk k^{1-2s} \partial_k \ln \Sigma_{\te},\adb\\
E^{\tm}(s) &=& -\frac{\hbar c\cos\pi s}{\pi} \mu^{2s} \sum_{l=1}^\infty \nu \int_0^\infty dk k^{1-2s} \partial_k \ln  \Sigma_{\tm},
\end{eqnarray}
where the integrand functions are given by Eqs. (\ref{sigmas}). The summations in these expressions begin with $l=1$ because for $l=0$ the electromagnetic modes (\ref{fields}) are zero.

The derivative of the regularized energy with respect to the distance $d$ ($E(s) = E^{\te}(s) + E^{\tm}(s)$) may be found by interchanging the derivative and summation with integration. In manifest form it reads
\begin{eqnarray}
\partial_d E(s)&=&-\frac{\hbar c\cos\pi s}{\pi} \mu^{2s}
\sum_{l=1}^\infty \nu \int_0^\infty dk k^{1-2s} \partial_k \left\{
\frac{k(1-\varepsilon)}{\sqrt{\varepsilon}}\left[ {\cal
G}_{\te}^{-1} + {\cal G}_{\tm}^{-1}\right] \right\},\nonumber
\end{eqnarray}
where
\begin{eqnarray*}
{\cal G}_{\te} &=& \frac{1}{\sqrt{\varepsilon}}\frac{\Phi'_{\te}}{\Phi_{\te}} - 
\frac{e_l'(z_\varepsilon)}{e_l(z_\varepsilon)} =
\frac{\Sigma_{\te}}{e_l(z_\varepsilon)\Phi_{\te}},\adb\\
{\cal G}_{\tm} &=& - \frac{ \frac{\Phi'_{\tm}}{\Phi_{T\tm}} -
\frac{1}{\sqrt{\varepsilon}}
\frac{e_l'(z_\varepsilon)}{e_l(z_\varepsilon)}}{\frac{\Phi'_{\tm}}{\Phi_{\tm}}
\frac{e_l'(z_\varepsilon)}{e_l(z_\varepsilon)} + \frac{\nu^2
-\frac{1}{4}}{z^2 \sqrt{\varepsilon}}} =
-\frac{\Sigma_{\tm}}{z^2\left[e'_l(z_\varepsilon)\Phi'_{\tm,z} + e_l(z_\varepsilon)\Phi_{\tm} \frac{\nu^2 -\frac{1}{4}}{z^2 \sqrt{\varepsilon}}\right]}.
\end{eqnarray*}

Let us consider now the rared media with $\varepsilon (i\omega) = 1 + 4\pi N \alpha (i\omega) + O(N^2)$, where $\alpha$ is polarizability of the atom and the density of the dielectric matter $N\to 0$. In this case the Casimir energy $E(s)$ is expressed in terms the energy per unit atom $E_a(s)$ by relation
\begin{equation}
E(s) = N \int_d^\infty E_a(s) 4\pi (R+r)^2 dr + O(N^2).
\end{equation} 
From this expression it follows that 
\begin{equation}
E_a(s) = -\lim_{N\to 0}\frac{\partial_d E(s)}{4\pi N (R+d)^2},
\end{equation}
and in manifest form we obtain the interaction energy per unit atom  
\begin{equation}
E_a(s) = -\frac{\hbar c\mu^{2s} \cos\pi s}{\pi (R+d)^2} 
\sum_{l=1}^\infty \nu \int_0^\infty dk k^{1-2s} \partial_k \left\{
\frac{k\alpha (i\omega)}{G_{\te}} + \frac{k\alpha (i\omega)}{G_{\tm}}\right\},
\end{equation}
where
\begin{eqnarray*}
G_{\te} &=& \frac{\Sigma_{\te}}{e_l(z)\Phi_{\te}} =
\frac{f_{\te}(ik)}{e_l(z)\Phi_{\te}},\adb\\
G_{\tm} &=& -\frac{\Sigma_{\tm}}{z^2\left[e'_l(z)\Phi'_{\tm,z}
+ e_l(z)\Phi_{\tm} \frac{\nu^2 -\frac{1}{4}}{z^2 }\right]}
= -\frac{f_{\tm}(ik)}{e'_l(z)\Phi'_{\tm,z} + e_l(z)\Phi_{\tm} \frac{\nu^2 -\frac{1}{4}}{z^2}}.
\end{eqnarray*}
With definitions of the functions $\Phi_{\te}$ and $\Phi_{\tm}$ we have the
following relations
\begin{eqnarray*}
 \Phi_{\te} = s_l(z) f_{\te}(ik) - \frac{\Omega}{k} s_l^2(x) e_l(z),\adb \\
 \Phi_{\tm} = s_l(z) f_{\tm}(ik) + \frac{\Omega}{k} {s'}_l^2(x) e_l(z).
\end{eqnarray*}
Taking into consideration these expressions we express above formulas in slightly different form,
\begin{eqnarray*}
G_{\te}^{-1} &=& e_l(z)s_l(z) - \frac{Q}{x} \frac{s_l^2(x)
e_l^2(z)}{f_{\te}(ik)},\adb\\
G_{\tm}^{-1} &=& -e'_l(z)s'_l(z) - e_l(z)s_l(z) \frac{\nu^2 - \frac{1}{4}}{z^2} -
\frac{Q}{x} \frac{1}{f_{\tm}(ik)} \left[ {s'}_l^2(x) {e'}_l^2(z) + {s'}_l^2(x)
{e}_l^2(z) \frac{\nu^2 - \frac{1}{4}}{z^2}\right],
\end{eqnarray*}
by separating the terms which have no dependence on the parameter $Q = \Omega R$. By virtue of the fact that the Casimir energy is zero for an atom in vacuum ($Q = 0$) without boundaries, we subtract the terms with $Q = 0$ and define the interaction energy by the following relation
\begin{equation}
E_{\Omega} = \lim_{s\to 0} \{E_a(s) -  \lim_{\Omega \to 0} E_a(s)\}.
\end{equation}
With this definition we integrate by part over $k$ and arrive with the final formula ($x=kR,\ z=kL$)
\begin{equation}
E_\Omega = -\frac{\hbar c\Omega}{\pi (R+d)^2} \sum_{l=1}^\infty \nu \int_0^\infty dk \alpha (i\omega) \left\{ \frac{s_l^2(x) e_l^2(z)}{f_{\te}(ik)} + \frac{{s'}_l^2(x) {e'}_l^2(z) + {s'}_l^2(x){e}_l^2(z) \frac{\nu^2 - \frac{1}{4}}{z^2}}{f_{\tm}(ik)} \right\},\label{Fcp}
\end{equation}
where the Jost functions on the imaginary axes read
\begin{eqnarray}
f_{\te}(ik) &=& 1 + \frac{\Omega}{k} s_l(x)e_l(x) ,\\
f_{\tm}(ik) &=& 1 - \frac{\Omega}{k} s'_l(x)e'_l(x).
\end{eqnarray}

To perform computations one needs an expression for the atomic dynamic polarizabilities of hydrogen. In was shown in Ref. \cite{JohEps67} that the precise expression for the atomic dynamic polarizability of hydrogen is given by the $10$-oscillator formula 
\begin{equation}
 \alpha (i\omega) = \sum_{k=1}^{10}\frac{g_{k,a}^2}{\omega^2 + \omega^2_{k,a}},
\end{equation}
where $g_{k,a}$ are the oscillator strengths and $\omega_{k,a}$ are the eigenfrequencies. All these parameters  may be found in Refs. \cite{JohEps67,BlaKliMos05}. It was shown in Ref. \cite{BlaKliMos05} that the polarizabilities can be represented with sufficient precision in the framework of the single-oscillator model  
\begin{equation}
 \alpha (i\omega) = \frac{g_a^2}{\omega^2 + \omega^2_a},\label{alpha}
\end{equation}
where $\alpha_a (0) = 4.50\ a.u.$ ($1\ a.u. = 1.482 \cdot 10^{-31} m^3$) and $\omega_a = 11.65 eV$ for hydrogen atom.  

One can see from the expression (\ref{Fcp}) that the energy is negative because the integrand is positive for arbitrary radius of the sphere, the wave number $\Omega$ and arbitrary position of atom. The same observation was noted in Ref. \cite{JheKim95} for ideal case. Let us consider different limits.

1) In the limit of perfect conductivity, $\Omega\to\infty$, which we call the Boyer limit, we obtain 
\begin{equation}
E_B = -\frac{\hbar c}{\pi (R+d)^2}\sum_{l=1}^\infty \nu \int_0^\infty dk k \alpha (i\omega)
\left\{ \frac{s_l^2(x) e_l^2(z)}{s_l(x)e_l(x)} - \frac{
{s'}_l^2(x) {e'}_l^2(z) + {s'}_l^2(x)
{e}_l^2(z) \frac{\nu^2 - \frac{1}{4}}{z^2}}{s'_l(x)e'_l(x)}
\right\}.\label{FcpBoyer}
\end{equation} 

2) The limit of infinite radius of sphere, $R\to \infty$, with fixed distance, $d$, between the surface of sphere and an atom requires more machinery. One cannot merely interchange the limit  and summation and integration in above expressions \Ref{Fcp} and \Ref{FcpBoyer} because in this case the integrand has no dependence on the $l$ and the series is divergent. Indeed, in the limit of infinite radius of sphere 
\begin{eqnarray*}
 2s_l(x) e_l(z)|_{R\to \infty} &=& +e^{-kd},\ 2s_l(x) e_l(x)|_{R\to \infty} =+1,\\
 2s'_l(x) e'_l(z)|_{R\to \infty} &=& -e^{-kd},\ 2s'_l(x) e'_l(x)|_{R\to \infty}
=-1,\\
2s'_l(x) e_l(z)|_{R\to \infty} &=& +e^{-kd},
\end{eqnarray*}
and the sum over $l$ is divergent,
\begin{equation}
E_{\Omega} = -\frac{\hbar c\Omega}{2\pi (R+d)^2} \sum_{l=1}^\infty \nu \int_0^\infty dk \alpha (i\omega)
\frac{e^{-2kd}}{1+\frac{\Omega}{2k}} \to \infty.
\end{equation} 

In order to obtain the correct expression for the energy in the limit $R\to\infty$ we change the variable of integration $k\to \nu k$  in Eqs. \Ref{Fcp} and \Ref{FcpBoyer}
\begin{eqnarray}
E_{\Omega} &=& -\frac{\hbar c\Omega}{\pi (R+d)^2} \sum_{l=1}^\infty \nu^2 \int_0^\infty dk \alpha (i\omega\nu)\left\{ \frac{s_l^2(\nu x) e_l^2(\nu z)}{f_{\te}(ik \nu)} + \frac{{s'}_l^2(\nu x) {e'}_l^2(\nu z) + {s'}_l^2(\nu x){e}_l^2(\nu z) \frac{1 - \frac{1}{4\nu^2}}{z^2}}{f_{\tm}(ik\nu)} \right\},\adb\\
E_B &=& -\frac{\hbar c}{\pi (R+d)^2} \sum_{l=1}^\infty \nu^3 \int_0^\infty kdk \alpha (i\omega\nu)\left\{ \frac{s_l^2(\nu x) e_l^2(\nu z)}{s_l(\nu x)e_l(\nu x)} - \frac{{s'}_l^2(\nu x) {e'}_l^2(\nu z) + {s'}_l^2(\nu x){e}_l^2(\nu z) \frac{1 - \frac{1}{4\nu^2}}{z^2}}{s'_l(\nu x)e'_l(\nu x)}
\right\},
\end{eqnarray}
and use the uniform expansion for Bessel functions (see Ref. \cite{AbrSte70}). We obtain the following expressions
\begin{eqnarray}
 E_{\Omega} &=& -\frac{\hbar c\Omega}{\pi (R+d)^2} \sum_{l=1}^\infty \nu^2 \int_0^\infty dk \alpha (i\omega\nu)
e^{-2\nu [\eta(z) - \eta(x)]} \left\{\frac{xzt(x)t(z)}{4w} +
\frac{1+t^2(z)}{4pxzt(x)t(z)} + \ldots\right\},\label{EAs}\\
E_B &=& -\frac{\hbar c}{\pi (R+d)^2} \sum_{l=1}^\infty \nu^3 \int_0^\infty dk k\alpha (i\omega\nu) e^{-2\nu [\eta(z) - \eta(x)]} \left\{\frac{zt(z)}2 +
\frac{1+t^2(z)}{2zt(z)} + \ldots\right\},\label{EBAs}
\end{eqnarray}
where $p = 1+\frac{Q}{2\nu x^2 t(x)}$, $w = 1 + \frac{Q t(x)}{2\nu}$, $t(x) =
1/\sqrt{1+x^2}$, $\eta (x) = \sqrt{1+x^2} + \ln \frac{x}{1+\sqrt{1+x^2}}$ and
$x=kR,\ z=kL = k(R+d)$. In the limit of $R\to\infty$, the integrands in above both  expressions have the same form and the main contribution to the energy comes from the first term of uniform expansion,
 \begin{equation}
E = -\lim_{R\to \infty} \frac{\hbar c g^2}{\pi c^2 (R+d)^2} \sum_{l=1}^\infty \nu^3 \int_0^\infty \frac{dy y}{y^2 \nu^2 + q^2}\frac{e^{-2\nu [\eta(u) - \eta(y)]}}{ut(u)},
\end{equation} 
where $u = y (1+d/R),\ q_a = k_a R$ and we changed variable $k\to y = kR$. Here the single-oscillator model for polarizability (\ref{alpha}) was taken into account.    

\begin{figure}[ht]
%\centerline{\epsfxsize=11truecm\epsfbox{2.eps}}
\includegraphics[width=11truecm]{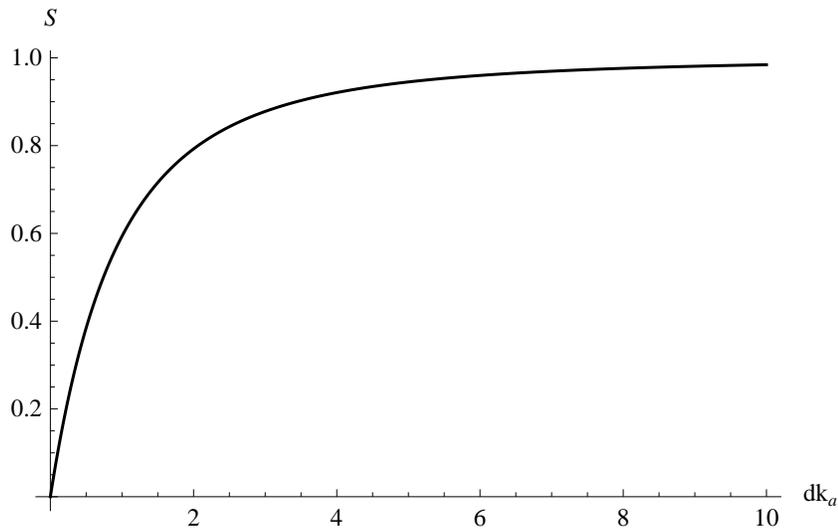}
\caption{\footnotesize The plot of $S$ as the function of the $v=kd_a$. It tends to unity for large $v$ ($E \sim d^{-4}$) and it is linear over $v$ ($E\sim d^{-3}$) for small distances between an atom and plate. The relation of the energy and $S$ is given by Eq. (\ref{defS}).}\label{fig:sboyerv}
\end{figure}

Next, the sum over $l$ we represent in the following integral 
\begin{equation}
\sum_{l=1}^\infty \frac{\nu^3 e^{-2\nu\delta}}{y^2 \nu^2 + q_a^2} = \frac{1}{4 q_a y} \int_{0}^\infty \frac{27 + 17 e^{-2(t+\delta)} + 5 e^{-4(t+\delta)}-e^{-6(t+\delta)}} {e^{3(t+\delta)}(e^{-2(t+\delta)}-1)^4 } \sin\frac{2q_at}{y} dt. 
\end{equation}
Assuming this expression we interchange the limit $R\to \infty$ and integrals over $y$ and $t$ and obtain 
\begin{equation}\label{defS}
E= -\frac{3\hbar c \alpha(0)}{8\pi d^4} S,
\end{equation}
where 
\begin{equation}
S = \frac 13 \int_0^\infty dt e^{-t} \left\{ \frac{1+t}{1+\frac{t^2}{4v^2}} + \frac{t}{(1+\frac{t^2}{4v^2})^2} \right\},
\end{equation}
and $v = dk_a$. Let us consider large distance, $d$, between the plate (sphere of infinite radius) and an atom, $d k_a \gg 1$. In the limit of $v\to\infty$ we obtain that $S =1$ and therefore the Casimir-Polder $(\sim d^{-4})$ energy, 
\begin{equation}
E= -\frac{3\hbar c \alpha(0)}{8\pi d^4},
\end{equation}
is recovered. For small distances, $dk_a \ll 1$, we change the variable $t\to \tau = t/2v$ and take the limit of $v \to 0$. In this case we obtain that $S = \pi v/3$ and the energy has the form $\sim d^{-3}$,  
\begin{equation}\label{CasPolSmall}
E= -\frac{\hbar c \alpha(0)k_a}{8 d^3},
\end{equation}
as should be the case. The plot of the $S$ as function of variable $v=dk_a$ is shown in Fig. \ref{fig:sboyerv}. 

3) Let us analyze the energy for large ($d\gg k_a^{-1}, d\gg R$) and small ($d\ll k_a^{-1}, d\ll R$) distances between the sphere and an atom for finite $\Omega$ and $R$. In the case of large distance, $d\to \infty$, of an atom from the shell we use  Eq. (\ref{Fcp}). We change integrand variable $k=y/d$, next take limit $d\to \infty$, and then we take the integral over $y$. The main contribution comes from the first term with $l=1$:
\bs
\begin{eqnarray}
E_\Omega &\approx&  -\frac{3\hbar c \alpha(0)}{8\pi d^4} S_\Omega,\adb \\
S_\Omega &=& \frac{R^{3}}{d^{3}} \left\{ \frac{7Q}{3(3+Q)} + \frac{46}{3} F(a)\right\},\adb \label{sdinf} \\
F(a) &=& \frac{8a^{2}}{23} \int_0^\infty \frac{y^{4} + 2 y^{3} + 5 y^{2} + 6 y + 3}{3y^{2} + 2a^{2}}e^{-2y}dy, 
\end{eqnarray}
\es
where $a^{2} = Q d^{2}/R^{2} = d^{2} \Omega /R$. The first term in above expression (\ref{sdinf}) comes from \TE\ mode and second -- from \TM\ polarization. The function $F$  increases monotonically from zero for small $a$ ($d^{2} \ll R/\Omega$) to unity  for large $a$ ($d^{2} \gg R/\Omega$). In the case of $a \ll 1$ the function $F(a) \approx 2\pi \sqrt{6} a/23$. Therefore, in the limit of $\Omega \to 0$, the energy $E_\Omega \to 0$ as should be the case. 

Assuming a finite conductivity, $\Omega \not = 0$, and  large distance $d \gg k_a^{-1}, d \gg R, d \gg \sqrt{R/\Omega}$  we obtain that 
\begin{equation*}
S_\Omega = \frac{R^3}{d^3} \left\{ \frac{7Q}{3(3+Q)} + \frac{46}3 \right\}
\end{equation*}
and  we arrive with expression  
\begin{equation}\label{Egreat}
E_\Omega \approx  -\frac{\hbar c \alpha (0) R^3}{8\pi  (3+Q)d^7} (53 Q + 138).
\end{equation} 
Taking into account the Casimir-Polder interaction energy of two atoms with polarizations $\alpha$ and $\alpha_f$,
\begin{equation}
E = - \frac{23}{4\pi} \frac{\hbar c \alpha(0) \alpha_f(0)}{d^7},
\end{equation}
we observe that the sphere with finite conductivity has static polarizability 
\begin{equation}\label{alpha_2}
\alpha_f = \frac{53 Q + 138}{46 Q + 138}R^3.
\end{equation} 

To analyze the energy for small distances we use the following representation for the series 
\begin{equation}
\sum_{l=1}^\infty \frac{\nu^2}{y^2\nu^2 +q^2} \frac{e^{-2\nu \delta}}{1+\frac{a}{\nu}} = -\frac{1}{4(q^2 + a^2 y^2)} \int_0^\infty \left\{ f^{(2)} e^{-2 a x} + \frac{y}{2 q} f^{(4)}\sin \frac{2qx}{y} + \frac{ay}{q} f^{(3)}\sin \frac{2qx}{y}\right\}, 
\end{equation}
where $f(x) = e^{-3 (\delta + x)}/(1-e^{-2 (\delta + x)})$. The first and second terms give the $d^3$ contribution and the last term gives contribution $\sim d$. Taking into account these expressions we obtain that  
\begin{equation}
E= -\frac{\hbar c \alpha(0)k_a}{8 d^3}
\end{equation}
as should be the case, because close to the sphere we observe flat surface.  

\section{Numericals}\label{Sec:Num}

We analyze numerically the following expression for the energy ($x=kR,\ z=k(R+d)$)
\begin{equation}
E_\Omega = -\frac{\hbar c\Omega}{\pi (R+d)^2} \sum_{l=1}^\infty \nu \int_0^\infty dk \alpha (i\omega)
\left\{ \frac{s_l^2(x) e_l^2(z)}{f_{\te}(ik)} + \frac{
{s'}_l^2(x) {e'}_l^2(z) + {s'}_l^2(x)
{e}_l^2(z) \frac{\nu^2 - \frac{1}{4}}{z^2}}{f_{\tm}(ik)} \right\},\label{FcpNum}
\end{equation}
where the Jost functions in imaginary axes read
\begin{eqnarray}
f_{\te}(ik) &=& 1 + \frac{\Omega}{k} s_l(x)e_l(x) ,\\
f_{\tm}(ik) &=& 1 - \frac{\Omega}{k} s'_l(x)e'_l(x),
\end{eqnarray}
and polarizability of atom  has the single-oscillatory form 
\begin{equation}
 \alpha (i\omega) = \frac{g_a^2}{\omega^2 + \omega^2_a}.
\end{equation}
 
In the Boyer limit $\Omega \to \infty$ we obtain
\begin{equation}
E_B = -\frac{\hbar c}{\pi (R+d)^2} \sum_{l=1}^\infty \nu \int_0^\infty dk k \alpha (i\omega)\left\{ \frac{s_l^2(x) e_l^2(z)}{s_l(x)e_l(x)} - \frac{
{s'}_l^2(x) {e'}_l^2(z) + {s'}_l^2(x)
{e}_l^2(z) \frac{\nu^2 - \frac{1}{4}}{z^2}}{s'_l(x)e'_l(x)} \right\}.\label{FcpBoyerNum}
\end{equation}

For simplicity we extract as a factor the Casimir-Polder expression for the interaction energy of an atom with plate,  
\begin{equation}
 E_{\Omega,B} = -\frac{3\hbar c \alpha (0)}{8\pi d^4} {S}_{\Omega,B},
\end{equation}
and we will numerically calculate the dimensionless quantities 
\begin{eqnarray}
{S}_\Omega &=& \frac{8 q_a^2 Q r^4}{3 (1+r)^2}\sum_{l=1}^\infty \nu \int_0^\infty  \frac{dy}{y^2 + q_a^2} \left\{ \frac{s_l^2(y) e_l^2(z)}{f_{\te}(iy)} + \frac{{s'}_l^2(y) {e'}_l^2(z) + {s'}_l^2(y) {e}_l^2(z) \frac{\nu^2 -\frac{1}{4}}{z^2}}{f_{\tm}(iy)} \right\},\\
{S}_B &=& \frac{8 q_a^2 r^4}{3 (1+r)^2}\sum_{l=1}^\infty \nu \int_0^\infty  \frac{y dy}{y^2 + q_a^2} \left\{ \frac{s_l^2(y) e_l^2(z)}{ s_l(y)e_l(y)} - \frac{{s'}_l^2(y) {e'}_l^2(z) + {s'}_l^2(y) {e}_l^2(z) \frac{\nu^2 -\frac{1}{4}}{z^2}}{s'_l(y)e'_l(y)} \right\},
\end{eqnarray}
where $z=(1+r)y$, $q_a = \omega_a R/c$, $r = d/R$,  $y = kR$ and
\begin{eqnarray}
f_{\te}(iy) &=& 1 + \frac{Q}{y} s_l(y)e_l(y) ,\\
f_{\tm}(iy) &=& 1 - \frac{Q}{y} s'_l(y)e'_l(y).
\end{eqnarray}

\begin{figure}[ht]
%\centerline{\epsfxsize=8truecm\epsfbox{3a.eps}\hspace*{2ex} 
\includegraphics[width=8truecm]{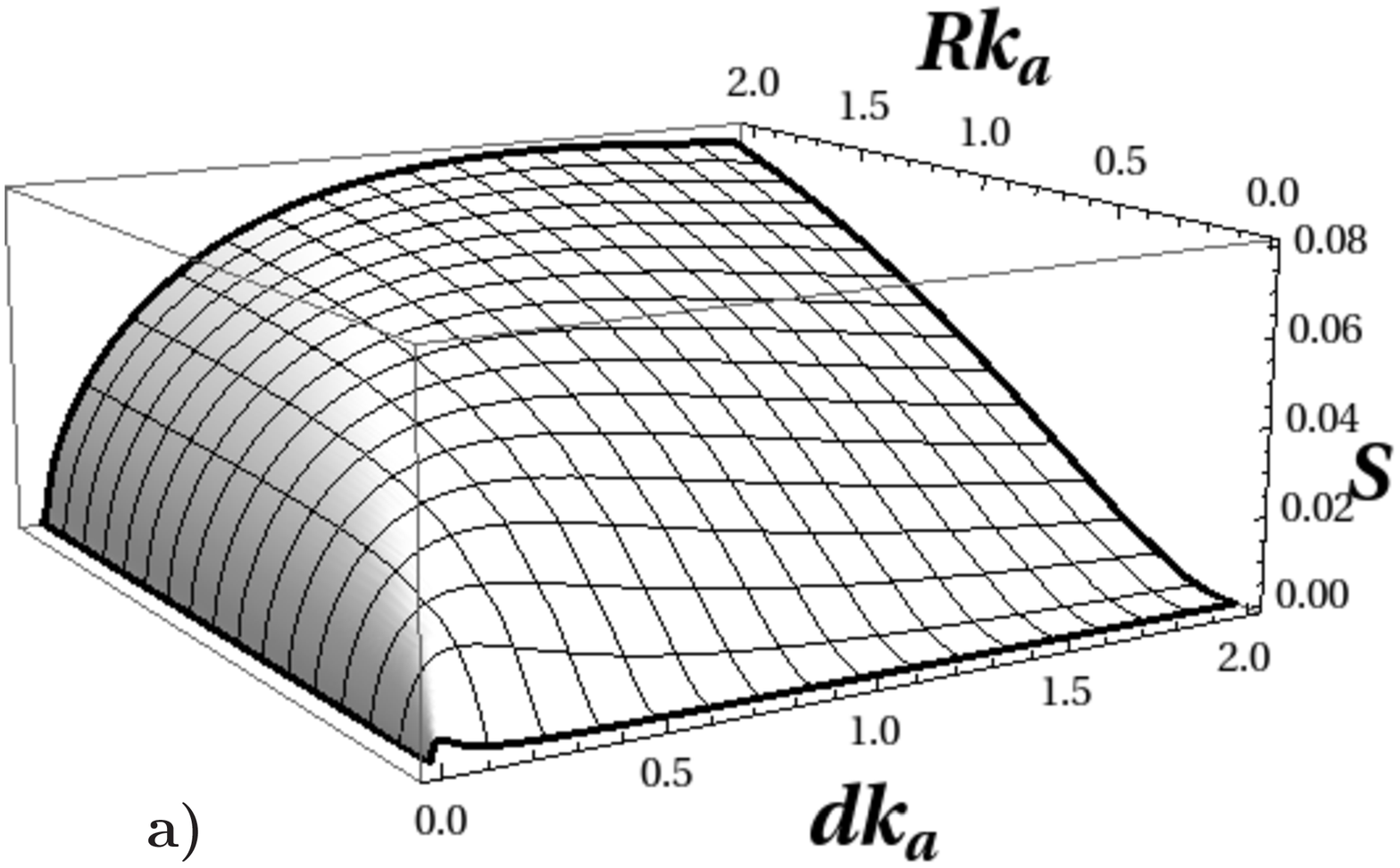}\hfill%
\includegraphics[width=8truecm]{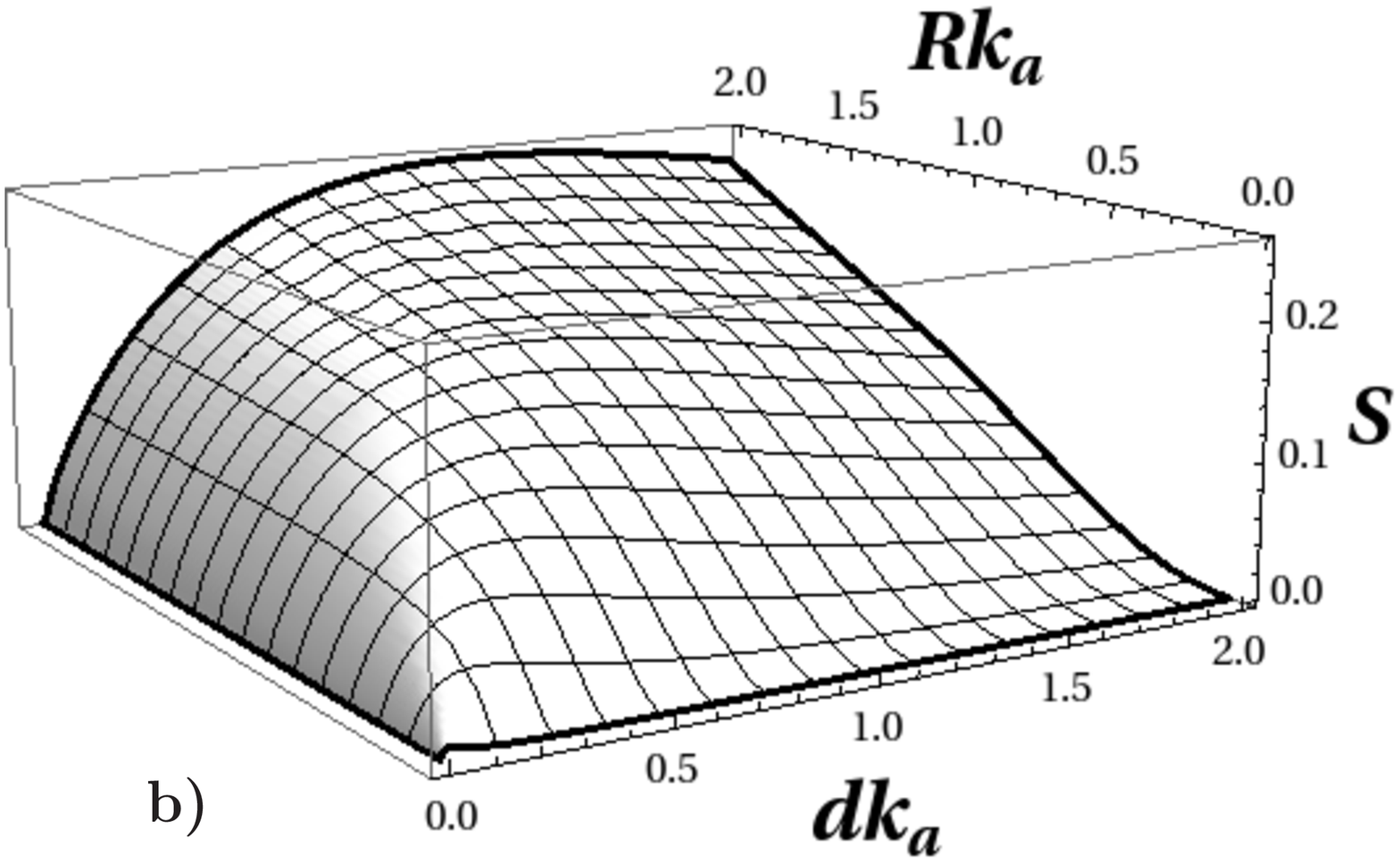}
\caption{\footnotesize The plot of $S$ as the function of the $dk_a\in (0,2)$ and $Rk_a\in (0.02,2)$ for $a)$ $\Omega/k_a = 2.44\cdot 10^{-2}$ and $b)$ $\Omega / k_a = 1$. }\label{fig:s1and2}
\end{figure}

We use $1/k_a$ as the unit of measurement of length and therefore the function $S$ depends on the three parameters: $\Omega/k_a, q_a=Rk_a$ and $ dk_a$. The numerical analysis of the function $S$ for $\Omega/k_a = 2.44\cdot 10^{-2}$ (molecule $C_{60}$) and $\Omega/k_a = 1$ is shown in Fig. \ref{fig:s1and2}.\vspace*{2em}

Let us consider the interaction energy between hydrogen atom and molecule $C_{60}$. For this molecule we have \cite{Bar04}: $R = 3.42 \textrm{\AA} = 0.342 nm$, $Q = \Omega R = 4.94 \cdot 10^{-4}$ and $\Omega/k_a = 2.44\cdot 10^{-2}$. The polarizability of hydrogen atom within the single-oscillator model reads  \cite{RauKleColBru82,BorGeyKliMos06,BlaKliMos07} $\alpha_a (0) = 4.50\ a.u.$ ($1\ a.u. = 1.482 \cdot 10^{-31} m^3$) and $\omega_a = 11.65 eV = 17.698 \cdot 10^{15} Hz\ (k_a = 0.059 nm^{-1},\ \lambda_a = 106.4 nm)$ where $\omega/c = k= 2\pi /\lambda$. Therefore, $q_a = k_a R = 0.0202$.

\begin{figure}[ht]
%\centerline{\epsfxsize=8truecm\epsfbox{4a.eps}
\includegraphics[width=8truecm]{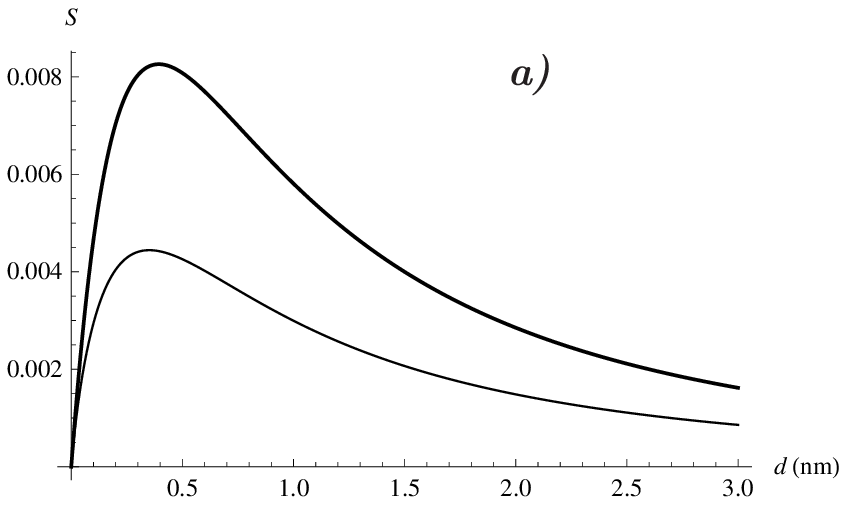}\hfill %
\includegraphics[width=8truecm]{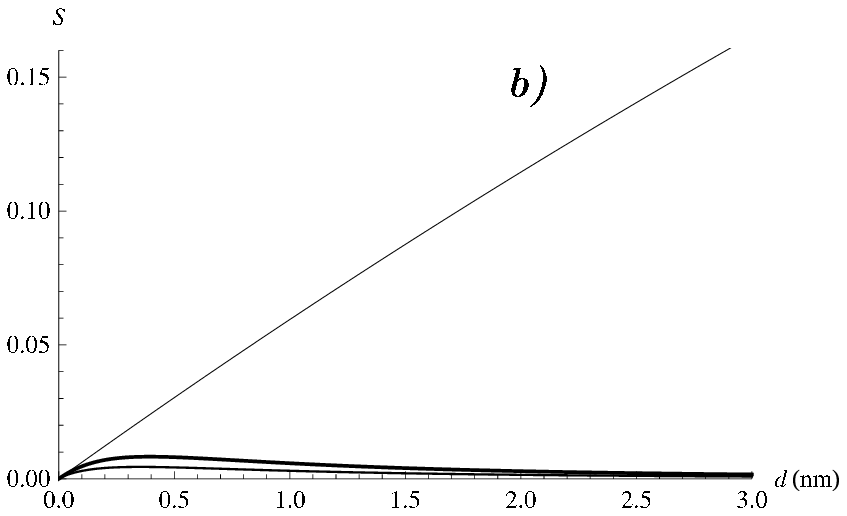}
\caption{\footnotesize The plot of $S$ as the function of the distance $d$ between an atom and the sphere. Thin curve is the energy for the case $R \to \infty$ (Casimir-Polder energy for plate), middle thickness curve is the case of the molecule $C_{60}$, and the thick curve is the case of ideal sphere ($\Omega \to \infty$). In the figure $b$ we compare the energy for the plane with the energy in the sphere case.}\label{fig:srd}
\end{figure}

\begin{figure}[ht]
%\centerline{\epsfxsize=8truecm\epsfbox{5a.eps}
\includegraphics[width=8truecm]{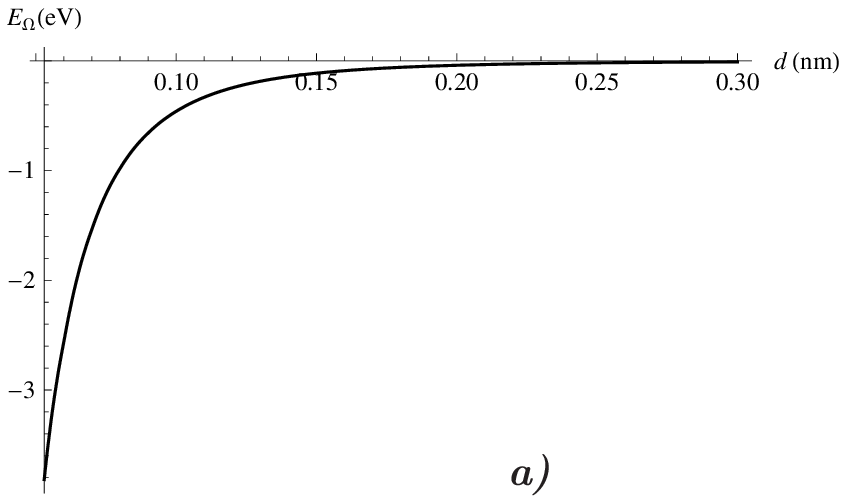}\hfill %
\includegraphics[width=8truecm]{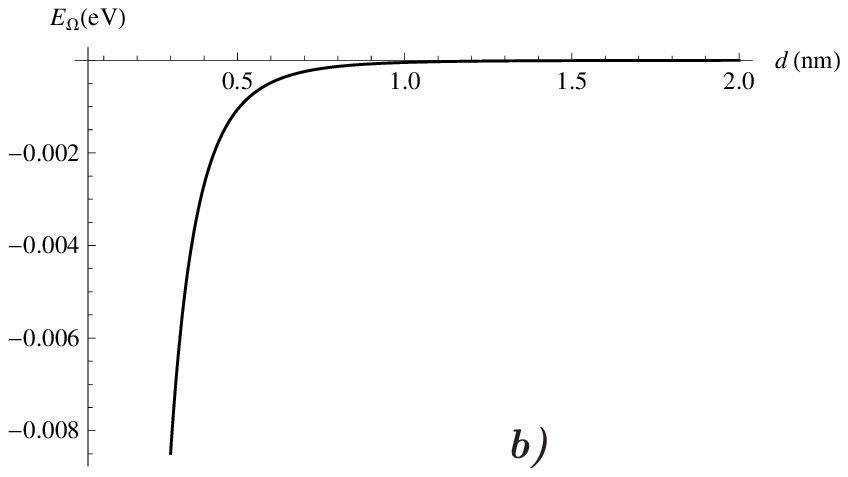}
\caption{\footnotesize The plot of the energy $E_\Omega$ as the function of the distance $d$ between the sphere and the hydrogen atom. In the figure $a)$ we show the energy starting from the distance $d = 0.053 (nm)$ (the radius of the hydrogen atom). In the  figure $b)$  the energy in large interval is shown.} \label{fig:srdhyd}
\end{figure}

Taking into consideration  all the numerical values of parameters we represent the energy for this system in the following form
\begin{equation}
 E_{\Omega}(eV) = -\frac{0.0156}{d^4(nm)} S_{\Omega}(q_a,r), 
\end{equation}
where the energy is measured in $eV$ and the distance is measured in nanometres. The numerical simulations for the function $S$ are shown in Fig. \ref{fig:srd} and the energy $E_\Omega$ in Fig. \ref{fig:srdhyd}. 
The radius of the hydrogen atom is $r_H = 0.053 nm$. For this minimal distance, $d=r_H$, we have numerically $E = 3.8 eV$. In the case of plate with hydrogen atom we obtain $6.4 eV$. In the interval of distances from the hydrogen atom radius $r_H$ up to $5r_H$ the energy is approximated by the following expression
\begin{equation}
E_\Omega(eV) \approx -\frac{0.00013}{d^{7/2}(nm)}. 
\end{equation}  
The same dependence was observed in Ref. \cite{ChuFedKliYur10}.

For large distances we obtain from Eq. (\ref{Egreat})
\begin{equation}\label{Efar}
E_\Omega(eV) \approx -\frac{0.0095}{d^7(nm)}.
\end{equation}
This expression approximates the exact one with error $10\%$ starting with distance $d=50nm$. The Eq. (\ref{alpha_2}) gives the static palarizability of the fullerene $\alpha_p(0) = R^3 = 4\cdot 10^{-29} m^3$. This expression is close to that calculated in Ref. \cite{FowLazZan91} where the authors obtained $\alpha_p(0) = 7\cdot 10^{-29} m^3$. 

\section{Conclusion} \label{Sec:Conc}

In the foregoing, we have obtained the analytic expression for the Casimir-Polder (van der Waals) energy for a system which contains an atom or microparticle and  infinitely thin sphere with finite conductivity which models a fullerene. We used the zeta-regularization approach and for renormalization we used a simple physically reasonable condition -- the energy should be zero for an  atom alone without a sphere. The conductive sphere with radius $R$ is characterized by the only parameter $\Omega = 4\pi n e^2/mc^2$ with dimension of wave number, where $n$ is the surface density of electrons. The limit $\Omega \to \infty$ corresponds to the ideal case considered by Boyer \cite{Boy68}. The microparticle is characterized by the only parameter, polarizability $\alpha$. 

The expression obtained reproduces in the limit $R\to \infty$ the Casimir-Polder result for an atom and plate (see Eqs. (\ref{defS})-(\ref{CasPolSmall})). For small distances we have $d^{-3}$ dependence and far from the plate we obtain $d^{-4}$ due to retardation. For finite radius of the sphere we have different behavior of the energy. Close to the sphere,  $d\ll 1/k_a$ and $d\ll R$, we have the same $d^{-3}$ dependence as in the Casimir-Polder case and far from the sphere we obtained $d^{-7}$ dependence given in Eq. (\ref{Egreat}). This expression is valid for $d\gg 1/k_a$ and $d\gg R$. For the interval $r_H < d < 5r_H$, where $r_H$ is the radius of the hydrogen atom, the energy is approximated by $d^{-7/2}$ dependence. We also note that the finite conductivity decreases the energy in comparison with Boyer case which may be observed in Fig. \ref{fig:srd}. 

Application to the molecule $C_{60}$ with hydrogen atom is plotted in Fig. \ref{fig:srdhyd}. For closest distance atom from the fullerene, which is radius of hydrogen atom $r_H$, the energy is $3.8 eV$ which is two times smaller then for the case of hydrogen atom with plate. Away from the fullerene (in fact larger then $50 nm$) the energy falls down as $d^{-7}$ (see Eq. (\ref{Efar})) which is in three orders of
magnitude faster then for the Casimir-Polder case. This dependence corresponds to the Casimir-Polder interaction atoms for large distance. Taking into account this analogy we obtain the polarizability of fullerene ($Q = \Omega R = 4.94\cdot 10^{-4} \ll 1$) 
\begin{equation*}
\alpha_f = \frac{53 Q + 138}{46 Q +138} R^3 \approx R^3 = 4\cdot 10^{-29} m^3.
\end{equation*} 
This expression is close to that calculated in Ref. \cite{FowLazZan91} where the authors obtained $\alpha_p(0) = 7\cdot 10^{-29} m^3$. 

In the paper we considered the interaction energy in the framework of the hydrodynamical model. As it was noted in the Introduction this model does not describe correctly graphene and therefore the systems made of them such as fullerenes. The model which describes graphene more precisely is the Dirac model. Nevertheless, using the calculations within the Dirac model which was made in Ref. \cite{ChuFedKliYur10}  as the base we expect that the interaction energy in framework of the Dirac model will be in five times smaller at large distance between fullerene and an atom. The dependence on the energy for large and small  distances between fullerene and an atom is expected to be the same. 

There is another question which was not considered in the paper but which is very important for condensed matter physics. It is interesting to obtain the adsorption energy of the hydrogen on the $C_{60}$ at the physical equilibrium distance. This question is very important for the problem of storage of hydrogen in carbon nano-systems (see Ref. \cite{DilJonBekKiaBetHeb97} and review \cite{Nec06}). We plan to investigate these questions in the future works.    

\acknowledgments 
The author would like to thank V. Mostepanenko and G. Klimchitskaya for stimulation of
these calculations and M. Bordag for discussions. This work was supported by the Russian Foundation for Basic Research Grant No. 08-02-00325-a.

\end{document}